\documentclass{iaus}
\usepackage{graphicx}

\title[Planetary Nebulae abundances] 
{Planetary nebulae in the inner Milky Way}

\author[O. Cavichia, R.D.D. Costa \& W.J. Maciel]   
{Oscar Cavichia$^{1,2}$
 , Roberto D.D. Costa$^1$   \and Walter J. Maciel$^1$}

\affiliation{$^1$IAG, University of S\~ao Paulo, 05508-900, S\~ao Paulo-SP, Brazil.\\
$^{2}$ email: cavichia@astro.iag.usp.br }

\pubyear{2009}
\volume{265}  
\pagerange{}
\jname{Chemical Abundances in the Universe: Connecting First Stars to Planets}
\editors{K. Cunha, M. Spite \& B. Barbuy, eds.}
\begin{document}

\maketitle

\begin{abstract}
 
New abundances of planetary nebulae located towards the bulge of the Galaxy are derived based on observations made at LNA (Brazil). We present accurate abundances of the elements He, N, S, O, Ar, and Ne for 56 PNe located towards the galactic bulge. The data shows a good agreement with other results in the literature, in the sense that the distribution of the abundances is similar to those works. From the statistical analysis performed, we can suggest a bulge-disk interface at 2.2 kpc for the intermediate mass population, marking therefore the outer border of the bulge and inner border of the disk.
\keywords{Planetary nebulae, chemical abundances,  chemical evolution, Milky Way}
\end{abstract}

\firstsection 
\section{Introduction}


Bulge and disk may have formed in different ways such as the disk inside-out formation model (Chiappini et al. \cite{chiap01}), and the model of multiple infalls onto the bulge (Costa et al. \cite{costa08}), so that we would expect that these differences should appear in the abundance distributions of these structures. Many authors have compared the abundance distributions of the bulge and the disk and find no clear differences (Chiappini et al. \cite{chiap09}, Escudero et al. \cite{escu04}, Exter et al. \cite{exter04}, Cuisinier et al. \cite{cuisin00}). Nevertheless, in these works abundances of the solar neighborhood or the whole disk were used in order to compare the abundance distributions. Until now, few studies made an effort to investigate whether or not  the radial abundance gradient of the disk extends toward the galactic center, as for example those from Smartt et al. (\cite{smartt01}) or Gutenkunst et al. (\cite{guten08}). In this context, the present study intends to shed light in this field by comparing the PNe abundance distributions of the inner disk and the bulge using PNe statistical distance scales. 

\section{Method}

Spectrophotometry observations in the optical domain were made at LNA observatory (Brazil) for a sample of 56 planetary nebulae located in the direction of the the galactic bulge.  The data were reduced following standard reduction procedures with the IRAF software (see Escudero et al. \cite{escu04} for details).

The Stanghellini et al. (\cite{stang08}) (SSV08) statistical distance scale was used to study the distribution of chemical abundances across the disk-bulge interface. Additionally, new distances were derived for 46 objects whose distances were not available in the same paper. The method consists in establishing a galactocentric distance that divides the sample into two groups: group I, composed by those PNe with distances lower than the limit, and group II with objects whose distances are higher than the limit settled. Then the galactocentric distance that divides the groups is varied from 0.1 to 3.6 kpc, in 0.7 kpc steps. A Kolmogorov-Smirnov test was then applied to each step in order to find the distance in which the chemical properties of these regions better separates.

\section{Results and discussion}

\begin{figure}[!ht]
\centering
\includegraphics[scale=0.55]{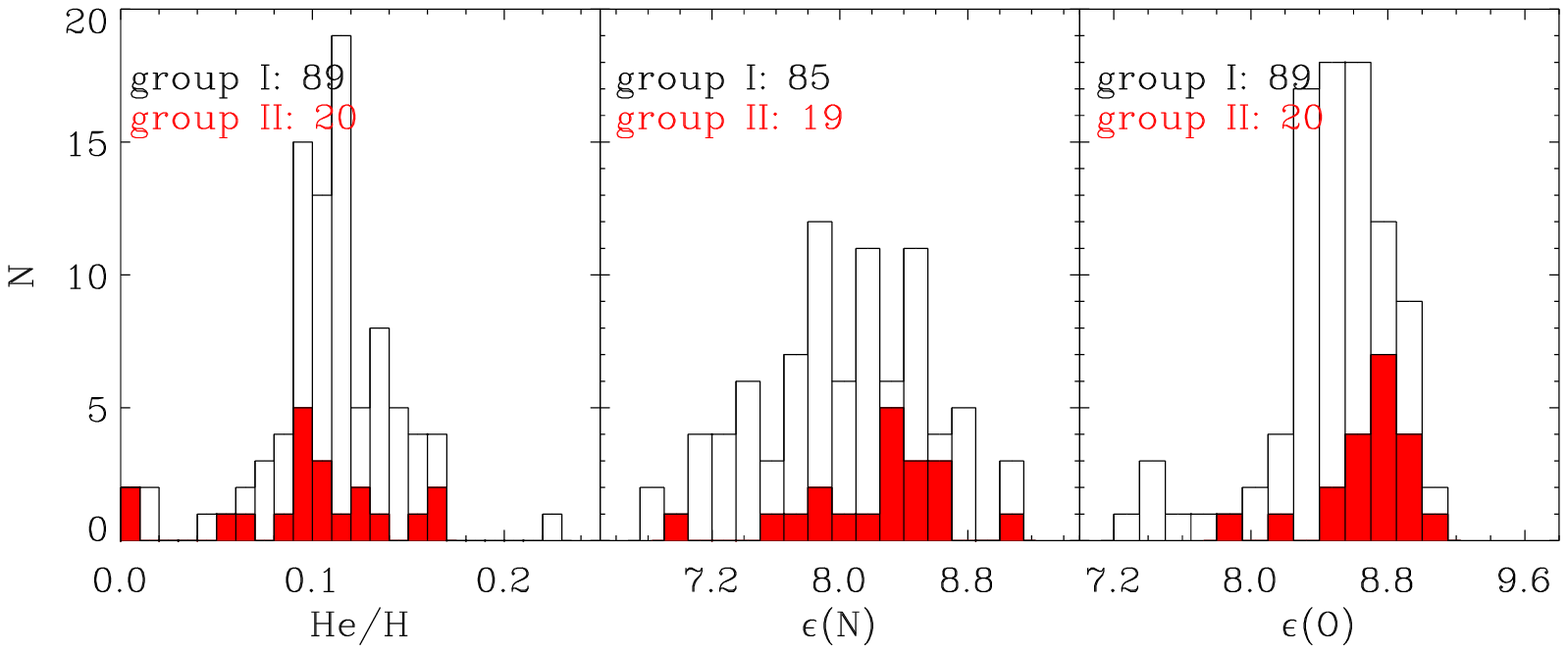}
\includegraphics[scale=0.55]{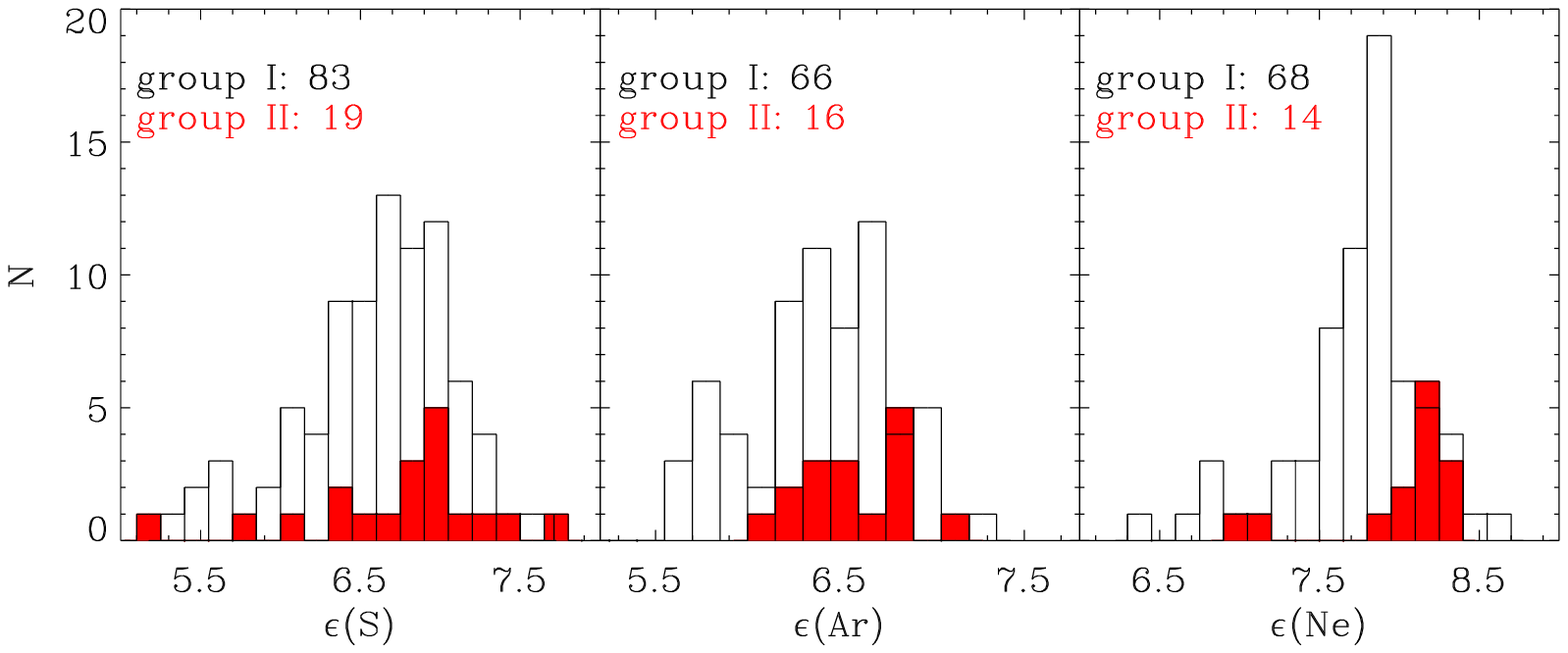}
\caption{\label{fig1} Abundance distributions for each chemical element for groups I and II using the SSV08 distance scale and a galactocentric distance for the separation set at 2.2 kpc. Unfilled histograms represent group I objects and filled histograms are for group II.  The number of objects in each distribution is shown at the top.}
\end{figure}

The Kolmogorov-Smirnov test results in a galactocentric distance of 2.2 kpc which better separates the two groups. Figure \ref{fig1} shows the abundance distributions for the two populations using this  distance. 

The comparison between the two populations shows that, on the average, group I (bulge) objects have slightly lower abundances than those from the group II (inner-disk), although this difference is not larger than the errors in individual abundances. Taking into account the results derived in this work as well as other evidences from the literature, and using the SSV08 distance scale, we propose a galactocentric distance of 2.2 kpc to mark the transition between the bulge and inner-disk of the Galaxy.\\

\textit{Acknowledgements.} This work was partially supported by FAPESP and CNPq.


\begin{thebibliography}{}

\bibitem[2001]{chiap01} Chiappini, C., Matteucci, F., \& Romano, D.\ 2001, \textit{ApJ}, 554, 1044 	 \bibitem[2009]{chiap09} Chiappini, C., Gorny, S., Stasi\'nska, G., \& Barbuy, B.\ 2009, \textit{A\&A}, 494, 591
\bibitem[2008]{costa08} Costa, R.~D.~D., Maciel, W.~J., \& Escudero, A.~V.\ 2008, Baltic Astronomy, 17, 321 
 \bibitem[2000]{cuisin00} Cuisinier, F., Maciel, W.~J., K{\"o}ppen, J., Acker, A., \& Stenholm, B.\ 2000, \textit{A\&A}, 353, 543 
\bibitem[2004]{escu04} Escudero, A.~V., Costa, R.~D.~D., \& Maciel, W.~J.\ 2004, \textit{A\&A}, 414, 211 
 \bibitem[2004]{exter04} Exter, K.~M., Barlow, M.~J., \& Walton, N.~A.\ 2004, \textit{MNRAS}, 349, 1291 
  \bibitem[2008]{guten08} Gutenkunst, S., Bernard-Salas, J., Pottasch, S.~R., et al. \ 2008, \textit{ApJ}, 680, 1206 
\bibitem[2001]{smartt01} Smartt, S.~J., Venn, K.~A., Dufton, P.~L., et al. \ 2001, \textit{A\&A}, 367, 86 
\bibitem[2008]{stang08} Stanghellini, L., Shaw, R.~A., \& Villaver, E.\ 2008, \textit{ApJ}, 689, 194 (SSV08)
\end{thebibliography}
\end{document}